\documentclass[conference]{IEEEtran}
\usepackage{cite}
\usepackage{times}
\usepackage{soul}
\usepackage{url}
\usepackage[dvipsnames]{xcolor}
\usepackage[hidelinks]{hyperref}
\usepackage[utf8]{inputenc}
\usepackage[small]{caption}
\usepackage{graphicx}
\usepackage{amsmath}
\usepackage{amsthm}
\usepackage{booktabs}
\usepackage{tabularx}
\usepackage{algorithm}
\usepackage{algorithmic}
\usepackage[switch]{lineno}
\usepackage{amsfonts}
\usepackage{adjustbox}

\def\BibTeX{{\rm B\kern-.05em{\sc i\kern-.025em b}\kern-.08em
    T\kern-.1667em\lower.7ex\hbox{E}\kern-.125emX}}
    
\newcolumntype{C}{>{\centering\arraybackslash}X}

\def\BibTeX{{\rm B\kern-.05em{\sc i\kern-.025em b}\kern-.08em
    T\kern-.1667em\lower.7ex\hbox{E}\kern-.125emX}}

\begin{document}

\title{Analysing Biomedical Knowledge Graphs using Prime Adjacency Matrices}

\author{\IEEEauthorblockN{Konstantinos Bougiatiotis\textsuperscript{1,2} and Georgios Paliouras\textsuperscript{1}}
\IEEEauthorblockA{
\textsuperscript{1}\textit{Institute of Informatics and Telecommunications, NCSR Demokritos, Athens, Greece} \\
\textsuperscript{2}\textit{Department of Informatics and Telecommunications, National and Kapodistrian University of Athens, Athens, Greece} \\
\textit{Email: \{bogas.ko, paliourg\}@iit.demokritos.gr,
kbogas@di.uoa.gr}\\
}}


\maketitle

\begin{abstract}
Most phenomena related to biomedical tasks are inherently complex, and in many cases, are expressed as signals on biomedical Knowledge Graphs (KGs). In this work, we introduce the use of a new representation framework, the Prime Adjacency Matrix (PAM) for biomedical KGs, which allows for very efficient network analysis. PAM utilizes prime numbers to enable representing the whole KG with a single adjacency matrix and the fast computation of multiple properties of the network. We illustrate the applicability of the framework in the biomedical domain by working on different biomedical knowledge graphs and by providing two case studies: one on drug-repurposing for COVID-19 and one on important metapath extraction. We show that we achieve better results than the original proposed workflows, using very simple methods that require no training, in considerably less time.

\end{abstract}


\section{Introduction}

In recent years, the availability of big data has created a movement to develop methods that leverage more holistic views, by simultaneously considering data sources and interactions between their elements\cite{zitnik2019machine}. To this end, biomedical Knowledge Graphs (KGs)\cite{nicholson2020constructing} are becoming increasingly popular for tasks such as personalized medicine\cite{ping2017individualized}, predictive diagnosis\cite{su2020survey}, and drug discovery\cite{bonner2022review}.

When working with these biomedical KGs, one challenge consists of dealing with the highly coupled nature of entities that may be involved in a diverse set of biological pathways, molecular functions, diseases etc\cite{himmelstein2017systematic}. Another challenge is that performing analysis and predictions on these KGs usually enforces the different methodologies to utilize multi-hop neighbourhoods (e.g. capturing long-range dependencies in message-passing graph convolutional networks\cite{wu2021representing}). These challenges highlight the need for a framework that will facilitate easy and fast calculations of representations, that also capture the convoluted interactions of the entities.

To this end, we introduce the use of the \textit{Prime Adjacency Matrix} (PAM)\cite{bougiatiotis2022efficient} representation for biomedical KGs. It was introduced to deal with multi-relational networks to allow for compact representations and fast calculations of multi-hop matrices, making it ideal for applications on biomedical KGs. Through the PAM matrix, we can represent a KG in a single adjacency matrix without loss of information. At its core, is the Fundamental Theorem of Arithmetic, which allows the unique decomposition of any given positive integer to its prime factors. Using this theorem, we can map each unique relation type in the KG to a distinct prime, and then we can construct the PAM capturing all the information about the connectivity of the original KG, without loss.

Having a single adjacency matrix for the whole KG allows us to leverage multiple tools from classical network analysis. For example, we can easily calculate the powers of this matrix and generate the multi-hop adjacency matrices for the whole graph. This can be done very efficiently and allows us to scale different methodologies to many large-scale KGs. Moreover, these PAM matrices contain rich structural information about the graph, which is readily available for modeling; simply by looking up the values of the matrices, we can infer properties of the KG.

In this work, we motivate the use of PAMs in the biomedical domain by performing two case studies, one focused on drug repurposing for COVID-19 and one focused on ``metapath'' extraction that is important for finding meaningful paths for reasoning in KGs\cite{liu2021neural}. In both cases, we leverage the KG representation as expressed through the PAMs without training any model on the downstream application, to highlight the simplicity and richness of the representations. Moreover, we demonstrate the efficiency of the methodology by applying the framework to a collection of commonly-used biomedical KGs, with diverse structural properties.

Overall, the main contributions of this work are the following:
\begin{itemize}
    \item We introduce the use of Prime Adjacency Matrices on biomedical knowledge graphs.
    \item We showcase the efficiency of the framework on multiple knowledge graphs and perform two case studies on downstream tasks, without the need for training a specialized model. 
    \item We make the code and other materials publicly available and outline avenues of future work to facilitate progress in the field\footnote{It can be accessed  from \href{https://github.com/kbogas/PAM_CBMS}{https://github.com/kbogas/PAM\_CBMS}.}.
\end{itemize}

The rest of the paper is structured as follows: 
Section~\ref{sec:methodology} introduces the basics of the PAM framework and also provides some related work on the use cases that we will address. In Section~\ref{sec:applications} we will delve into the two use-cases, namely drug-repurposing for COVID-19 and metapath extraction. Finally, in Section~\ref{sec:conclusions}, we summarize the main aspects of the methodology employed and propose future work.




\section{Background and Related Work}\label{sec:methodology}
In this section, we introduce the  PAM framework, highlight its main features, and then provide related work on the downstream tasks we will address.

\subsection{Prime Adjacency Matrix (PAM)}
Let us start with a  directed, multi-relational knowledge graph $G=\{V,R,E\}$, with $V$ denoting the nodes of the graph (usually of different types), $R$ the distinct relation types and $E$ the set of edges in the graph. These edges are essentially triples of the form $\{(h, r, t):$ $h,t \in V, r \in R\}$, denoting that the node $h$ is connected to the node $t$, through a relation $r$.

We first associate each unique relation type $r \in R$ with a distinct prime number $p_r$. This is a simple procedure and in its simplest form, we would randomly order the relations and allocate the prime number 2 to the first relation, the prime number 3 to the second one, and so forth with the rest of the prime numbers until all distinct relations were mapped to a number.

With this mapping in place, $r \leftrightarrow p_r, \forall r \in R$, we can construct the \textit{Prime Adjacency Matrix} (PAM) for this graph $P$, with shape $|V| \times |V|$, in the following form:
\begin{equation}\label{eq:PAM}
        P[i,j]=
        \begin{cases}
             \displaystyle \mathop{\prod}_{r:(h,r,t) \in E} p_r & if\text{ $\exists r :$ $(h,r,t) \in E $} \vspace{0.3em}\\ 
             \hspace{2em} $0$ &if\text{ $\forall r :$ $(h,r,t) \notin E $}
        \end{cases}
\end{equation}

As we can see in \eqref{eq:PAM}, each non-zero element $P[i,j]$ is the product of the primes $p_r$ for all relations $r$ that connect node $i$ to $j$. The Fundamental Theorem of Arithmetic (FTA) states that we can decompose each product to the original primes that constitute it (i.e. the distinct relations that connect the two nodes), thus preserving the full structure of $G$ in $P$ without any loss.

\subsubsection{A simple example}

\begin{figure}[htbp]

\centerline{\includegraphics[width=0.98\linewidth]{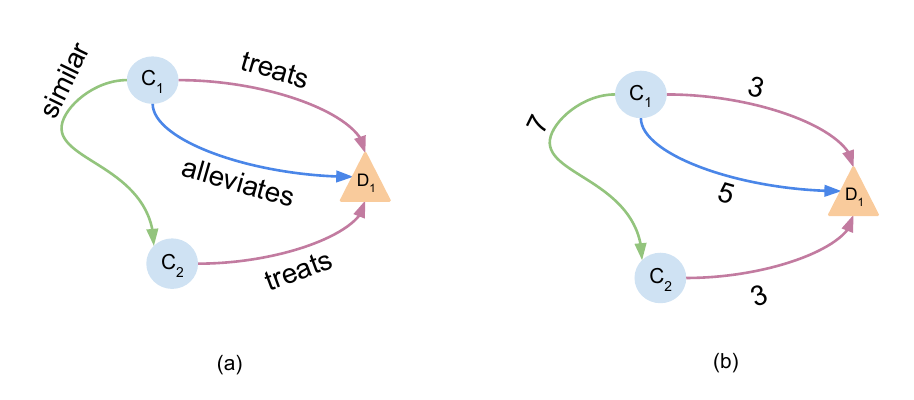}}

\caption{(a) An example multi-relational graph with 3 nodes (2 compounds and 1 disease) and 3 types of relation. (b) The same graph with the relations mapped to prime numbers.}
\label{fig:PAM_rolling_example}
\end{figure}

Let us consider a simple network, as the one shown in Fig.~\ref{fig:PAM_rolling_example}~(a), where we have 3 nodes (i.e. two compounds $C_1$, $C_2$ and a disease $D_1$) and 3 types of relation \textit{treats} (purple), \textit{alleviates} (blue) and \textit{similar} (green). The first step is to simply map each distinct relation to a prime number. Here the mapping is: \textit{treats} $\leftrightarrow 3$, \textit{alleviates} $\leftrightarrow 5$ and \textit{similar} $\leftrightarrow 7$ as shown in Fig.~\ref{fig:PAM_rolling_example}~(b).

The resulting PAM according to Eq.~\eqref{eq:PAM} would be (with node $C_1$ corresponding to index 0 of the matrix, node $C_2$ to index 1, and so forth):

\vspace{0.06cm}$P=
(\begin{smallmatrix}
0 & 7 & 15\\
0 & 0 & 3 \\
0 & 0 & 0 \\
\end{smallmatrix})$.\vspace{0.06cm} 

We see that the edge $C_1\xrightarrow{7}C_2$ is denoted by $P[0,1] = 7$ and $C_2\xrightarrow{3}D_1$ by $P[1,2] = 3$. Finally, we have the case where $C_1$ is connected to $D_1$ through the two distinct relations \textit{treats} and \textit{alleviates}, which are mapped to the primes $3$ and $5$ correspondingly. Thus, according to the definition of PAM, we have $P[C_1,D_1] = P[0,2] = 3 * 5 = 15$, which can be uniquely decomposed back to its prime factors, $3,5$, retrieving the original edges without loss.

Even in this toy graph, the compact PAM representation facilitates interesting observations. For example, we can see all the incoming/outgoing edges and their types by simply looking at the corresponding columns/rows of $P$. So, looking at $P[2,:]$ and $P[:,2]$ we see that node $D_1$ is a sink (i.e. has no outgoing edges of any kind). Another graph property that can be easily inferred is the frequency of different relations. If we simply count the frequencies of the prime factors of the non-zero elements of $P$, we get the distribution of edges per relation type, which is $\{3:2, 5:1, 7:1\}$.

\subsubsection{Moving to multi-hop matrices}

Having a single adjacency matrix for the whole knowledge graph allows us to utilize tools from classical network analysis. Most importantly, we can easily obtain the powers of the adjacency matrix.  In a single-relational network,  the element $(i, j)$ of the power $k$ of an adjacency matrix, contains the number of paths of length $k$ from node $i$ to node $j$. Generalizing this property to the PAM representation, where each value in the matrix also represents a specific type of the relation, the values of $P^k[i,j]$ allow us to keep track of the relational chains linking two nodes. 

For instance, the second-order PAM for the example graph of Fig.~\ref{fig:PAM_rolling_example} will be:\newline
$P^{2}= P\times P =
(\begin{smallmatrix}
0 & 0 & 21\\
0 & 0 & 0 \\
0 & 0 & 0  \\
\end{smallmatrix})$.\vspace{0.06cm}

For this toy example, the only non-zero element is $P^2[0,2] = P^2[C_1, D_1] = 21$ denoting a 2-hop path between $C_1$ and $D_1$. We can see from Fig.~\ref{fig:PAM_rolling_example}, that we can get from $C_1$ to node $D_1$ in two hops by following the directed path $C_1\xrightarrow{7}C_2\xrightarrow{3}D_1$. The relations $3$ and $7$ that are exhibited along this 2-hop path, are directly expressed in the value of $P^2[C_1,D_1]=21$, through its prime factors, as $21=3*7$.  Hence, using this representation, the products in $P^k$ express the relational $k$-chains linking two nodes in the graph.

Moreover, we can easily extract structural characteristics for nodes, pairs, subgraphs, and the whole graph, simply by looking up the PAMs. These can be used for further analysis according to the task at hand. For example, in this toy graph, we know that $C_1$ and $D_1$ are connected through the $2$-hop value $21$. This value semantically expresses the path: a compound ($C_1$) is similar to a second compound ($C_2$), which treats a disease ($D_1$). However, because we know that $C_1$ directly \textit{treats} $D1$, we could support the idea that this semantic pattern expressed by the value $21$, implies that the notion of \textit{treats} (i.e. $21\implies5$, meaning that if there exists a path between two nodes of the form $\{$\textit{similar},\textit{treats}$\}$, then these two nodes may also be directly connected through a \textit{treats} relation).

This procedure can be iterated for as many hops as we are interested in, by simply calculating the corresponding $P^k$. The values in each of these matrices will contain aggregated information regarding the relational chains of length $k$ that connect the corresponding nodes. Interesting characteristics about these graphs and their components can then be easily extracted through simple operations.

\subsection{Related Work}

\subsubsection{Drug Repurposing}

Drug repurposing is a strategy used for finding a drug from approved drugs to treat a specific disease (i.e. identifying new uses of known drugs), and it is implemented by identifying associations between drugs and diseases or inferring interactions between drugs and targets. Knowledge graphs are introduced to the domain of drug discovery for imposing an explicit structure to integrate heterogeneous biomedical data. The drug repurposing task is then formulated as knowledge graph completion, i.e. predicting unseen relations between two existing entities or predicting a candidate disease node given a drug node\cite{maclean2021knowledge}. A plethora of methodologies for tackling this problem have been proposed, from graph embeddings~\cite{GAO2022104133}, to neural~\cite{pan2022deep} and neuro-symbolic\cite{liu2021neural}. An overview of the different approaches and data sources used in this domain can be found in~\cite{ZENG2022114}. It is important to note that interest in these approaches was greatly raised due to the coordinated search for drug repurposing in the context of COVID-19\cite{al2021knowledge}.


\subsubsection{Metapath Extraction}

Drug repurposing can accelerate the process of drug discovery, however the main practical concern when using computational drug repurposing is the explainability aspect of it, which may hinder adoption in clinical settings~\cite{edwards2021explainable}. Most methods complement their prediction with intuitive explanations leveraging the semantically-rich KG-based paths that connect drug-disease pairs. For example, in~\cite{Ma2022.11.29.518441} they want to extract ``mechanism of action'' paths (which are essentially the paths on the knowledge graphs), that can semantically describe an abstract biological process of how a drug treats a disease. In \cite{sosa2019literature} they use an embedding model to score the edges in the graph and provide explanations for predictions via the highest-ranking paths based on confidence scores. On a similar basis, in~\cite{BAKAL2018189} the authors find important ``semantic patterns'' that could imply a hidden connection between drugs and diseases, based on the semantic relations exhibited along the paths connecting the drug-disease pairs, as extracted from the literature. In this study, we view these important multi-hop paths under the term of ``metapaths'' following the notation of \cite{himmelstein2017systematic}, as we will focus on the same KG as theirs, in our application.



\section{Applications}\label{sec:applications}

In the following subsections, we will present applications using the PAM representation. All experiments were run on an Ubuntu Server with Intel Core i7 Quad-Core @ $2.30$GHz. At most, $8$ threads and $16$ GB RAM were reserved for the experiments. 

\subsection{Calculating Prime Adjacency Matrices}
To showcase the usability of the PAM representation and the simplicity of the calculations needed, we aggregated some commonly-used biomedical knowledge graphs and generated their $P^k$ matrices. We selected knowledge graphs of varying properties (i.e. number of nodes, edges, relations, connectivity, etc.) in order to showcase the applicability of the framework.

The smaller (in terms of the number of edges) of the KG used is DDB14\cite{10.1145/3447548.3467247} which is a medical database containing biomedical entities 
and their relationships.\footnote{It was created based on DiseaseBank \url{http://www.diseasedatabase.com/}.}. Then, we have PharmKG\cite{10.1093/bib/bbaa344} a multi-relational, attributed biomedical KG, composed of more than 500,000 interconnections between genes, drugs and diseases, which is used mainly for data mining. Then, we have HetioNet~\cite{himmelstein2017systematic} which is a biomedical KG with multiple node and relation types, exhibiting also many hub nodes. DRKG\cite{drkg2020} is a COVID-specific knowledge graph created from aggregating databases and data from publications for the purpose of drug-repurposing.
The largest graph we experimented on is PrimeKG\cite{chandak2022building}, which is a recent KG that presents a holistic view of diseases by integrating more than 20 high-quality biomedical resources representing ten major biological scales.


\begin{table}[htbp]
\caption{Main characteristics of KGs and time needed to calculate PAMs up to $k=4$. Sp \% (k) denotes the percentage sparsity of $P^k$. \label{tab:scalability}}
\begin{center}
\adjustbox{max width=1\linewidth}{%
\begin{tabular}{lcccccc}

\toprule
 Dataset & $|V|$ &  $|R|$ & $|E|$ & 
Sp \% (1) & Sp \% (4) &  Total Time  \\
\midrule
   
   DDB14 &   9,057 &   14 &   36,561 &  99.96 &   99.22 &      0.20 (sec)     \\
 PharmKG & 188,296 &   39 & 1,093,236 & 100.00 &   92.44 &    5.24 (min) \\
Hetionet &  45,158 &   24 & 2,250,197 &  99.90 &   76.96 &     1.32 (min)  \\
    DRKG &  97,238&  107 & 5,874,260 &  99.95 &   73.24 &   11.95 (min)  \\
 PrimeKG & 129,262 &   30 & 8,100,498 &  99.95 &   79.38  &  11.40 (min) \\
\bottomrule
\end{tabular}
}

\end{center}

\end{table}

The basic characteristics of the six KGs can be seen in Table~\ref{tab:scalability}, where the columns represent: the number of nodes, the number of unique types of relations, the total number of edges, the sparsity of the PAM (in percentage) at the $1$- and $4$-hops and the total time taken to set up and calculate the PAMS up to $P^4$. The datasets are sorted from small to large, based on the total number of edges in each graph. We can see that for the case of DDB14, which is a small-scale KG, the whole process takes less than a second. For the rest of the medium- and large-scale KGs the whole process takes a few minutes\footnote{We stopped the calculations at $k=4$, because most KGs started exhibiting increasingly smaller sparsities, denoting an uninformative blur of connectivity between the nodes (most had less than $70\%$ sparsity).}.

Interestingly, the time needed to calculate $P^4$ for PharmKG is disproportionately longer than for Hetionet, which has double the number of edges. This is mainly due to the different structures of the KGs and the great number of nodes in PharmKG. It is important to note that we have not optimized the calculations of the PAMs, opting for simple sparse matrix multiplications between them. More efficient ways can be used to handle large-scale datasets~\cite{7013051}. It is also worth noting that, this procedure needs to be executed only once to calculate the needed $P^k$ and then they can be used for multiple downstream tasks. This means that with a few minutes of calculation, we obtain the higher-order associations between nodes, which can reveal valuable patterns for the task at hand.

\subsection{Drug Repurposing Study}

To showcase the applicability of the framework, we will use it to do a drug repurposing study for COVID-19 in a simple manner without the need for training a dedicated model. To start with, we will be focusing on DRKG\cite{drkg2020}, a Drug Repurposing Knowledge Graph for COVID-19, created from structured databases and publications related to COVID-19 (details on DRKG can be seen in Table~\ref{tab:scalability}). Having constructed this KG, the creators in ~\cite{drkg2020} devised a simple procedure to find possible drugs for the disease\footnote{As also described \href{https://github.com/gnn4dr/DRKG/blob/master/drug_repurpose/COVID-19_drug_repurposing.ipynb}{here}.}. Specifically, they used $8140$ drugs from DrugBank as candidates, that are FDA-approved and with molecular weight more than 250 daltons. Moreover, they use $34$ entities in the KGs as representatives of the target disease (e.g. \textit{Disease::SARS-CoV2 Spike}, \textit{Disease::MESH:D045169}), which are essentially variants or duplicates of the disease from different databases. Thus, they want to predict possible treatment links between these drug nodes and the disease nodes. 

In order to do that, they train a graph embedding model (specifically TransE\cite{bordes2013translating}), in order to generate embeddings for the nodes and the relations. Having this model in place, they then score all $8140\times34$ combinations of drugs and diseases and rank them from the most probable to the least one. In order to evaluate them, they test how many of the top-100  drugs have been also tested in a clinical trial\footnote{The clinical trial drugs that were mentioned at the time were 32 and taken from \url{http://www.covid19-trials.com/}, as accessed on September'21.}.
They found 5 candidates in their top-100 that were also used in some clinical trial\footnote{Excluding \textit{Ribavirin} which was already included as a treatment for SARS in the KG.}, and these are shown in Table~\ref{tab:dr_results}.

\begin{table}[htbp]
\caption{Predicted candidate drugs in the top-100 for COVID-19, that were also used in a clinical trial. The number in parenthesis denotes the rank  of the candidate in the top-100  results (lower is more probable).\label{tab:dr_results}}
\begin{center}
\adjustbox{max width=1\linewidth}{%
\begin{tabular}{cc}

\toprule
 \textit{TransE} & \textit{PAM-SVD}  \\
\midrule
Dexamethasone (4) & Dexamethasone (1) \\
Colchicine (8) & Methylprednisolone (4) \\
Methylprednisolone (16) & Colchicine (27)  \\
Oseltamivir (49)& Thalidomide (33) \\
Deferoxamine (87)& Deferoxamine (50)  \\
-& Azithromycin (57) \\
-& Oseltamivir (59) \\
-&  Chloroquine (69)\\
-&  Hydroxychloroquine (90)\\
\bottomrule
\end{tabular}
}
\end{center}

\end{table}

Our task is to explore whether the PAM representation makes it easy for us to identify these candidate drugs. To do that, we first construct the $1$-hop PAM matrix of DRKG, according to Eq.\eqref{eq:PAM}. Then as a very simple predictor, we utilize a matrix factorization method that is commonly used for link prediction\cite{menon2011link}. Specifically, we approximate the original matrix $P$ through $\tilde{P}=U * S * V$, a low $k$-rank approximation of the matrix, where we  only use the $k$ first eigenvalues of the eigendecomposition of $P$ to approximate the original matrix. For our purpose, we chose $k=200$ to be on the same scale as the hidden embedding size of $h=400$ of the TransE model.

Using this approximation, we score all the combinations of drugs and diseases (i.e. we look up the cell values for these pairs in $\tilde{P}$) and rank them from highest to lowest. Using this simple procedure, we found 9 candidates in the top-100 highest-scoring ones, including all those found using the TransE embeddings, as shown in the \textit{PAM-SVD} column of Table~\ref{tab:dr_results}. We can see that 3/5 of the drugs proposed by TransE were scored higher through our model. Moreover, if we use as golden truth drugs, the ones that have been used in more recent clinical trials\footnote{As the ones found in \url{https://go.drugbank.com/covid-19}, where 708 drugs are used in clinical trials.}, the TransE method has 32 hits at its top-100, while the unsupervised \textit{PAM-SVD} method has $45$ hits, again outperforming the trained embeddings model. Also, let us note that the \textit{PAM-SVD} procedure takes less than two minutes in total in a conventional laptop, for constructing the PAM matrix and calculating the low-rank approximation, while in comparison, the TransE embeddings are trained for multiple epochs until convergence (the authors mention training on an AWS machine with 8 GPUs).

Overall, we showcased the efficiency and simplicity of using PAM on the task of drug repurposing. As shown, for a medium-scale heterogeneous knowledge graph as DRKG we can have the PAM representation constructed very fast and then perform numerous downstream tasks on top. Moreover, having a single matrix to represent the whole KG in a compact form allows us to leverage well-studied techniques from classical network analysis, such as the low-rank approximation used here. Finally, it is also worth noting that in the above procedure, no training of any model was required, allowing for fast prototyping of ideas.

\subsection{Metapath Extraction Study}\label{subsec:MetaPath}
As discussed previously, through the PAM representation we can easily calculate the multi-hop paths between two nodes using the values of the $P^k$ matrices. These multi-hop paths express the chain of relations that connect two nodes in the graph. These ``metapaths'' capture semantics that express specific relationships between the entities and can also be used to hypothesize and provide explanations for possible links that may connect these entities. To showcase this we will be focusing on  HetioNet~\cite{himmelstein2017systematic}, which was created by integrating more than 29 resources containing diseases, genes, anatomies, pathways, compounds, disease symptoms, and many more. In \cite{himmelstein2017systematic}, they devised a methodology to find metapaths in this heterogeneous graph that captures patterns that can be used to identify specific chemical compounds (i.e. drugs) suitable for treating a disease. These metapaths express naturally some mechanisms of pharmacological efficacy. For example, a metapath of the form Compound–\textit{binds}–Gene–\textit{associates}–Disease (CbGaD) identifies when a drug binds to a protein, corresponding to a gene, involved in the disease. 

To find these metapaths, they used the existing Compound-\textit{treats}-Disease relations as supervision labels to train machine learning models and extract important metapaths. Through their methodology, they came up with various important metapaths. Those with a length of $k=2$ are shown in Table~\ref{tab:metapaths_hetionet}.

    \begin{table}[htbp]
    \caption{Important metapaths for drug efficacy.}
    \begin{center}
    \begin{tabular}{lc}
    \hline
    \textbf{Metapath} & \textbf{Acronym}\\
    \hline
    Compound–\textit{resembles}–Compound–\textit{treats}–Disease  & CrCtD \\
    Compound–\textit{treats}–Disease–\textit{resembles}–Disease & CtDrD \\
    Compound–\textit{binds}–Gene–\textit{associates}–Disease & CbGaD\\
    Compound–\textit{downregulates}–Gene–\textit{upregulates}–Disease & CdGuD \\
    Compound–\textit{upregulates}–Gene–\textit{downregulates}–Disease & CuGdD \\
    \hline
    \end{tabular}
    \label{tab:metapaths_hetionet}
    \end{center}
    \end{table}
    
\begin{figure}[htbp]

    \centerline{\includegraphics[width=0.5\textwidth]{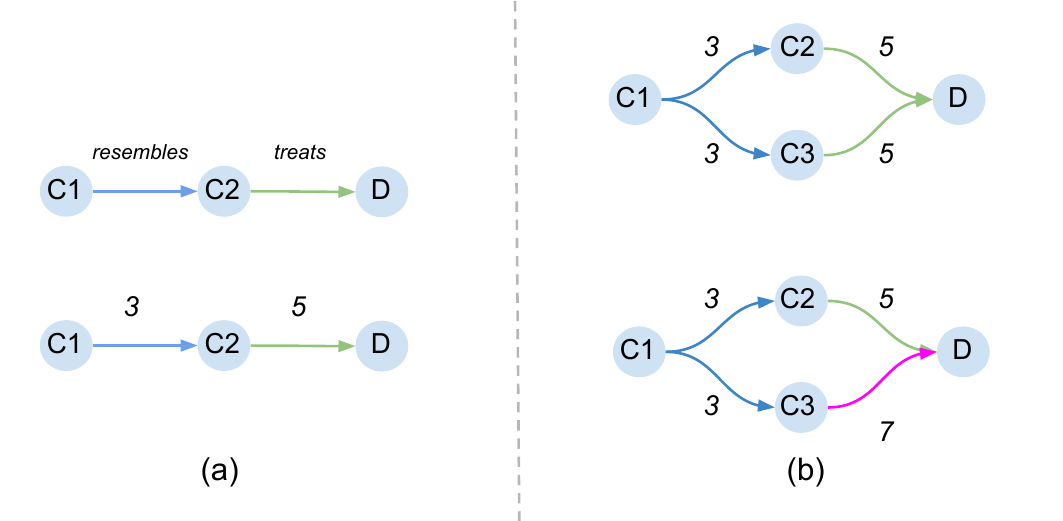}}

    \caption{Example of metapath patterns in HetioNet. (a) The metapath CrCtD is mapped to $3*5=15$. (b) Examples of aggregates of paths connecting a compound and a disease.}
    \label{fig:PAM_metapaths}
    \end{figure}

Our task is to explore whether we can easily identify these metapaths using PAMs. To do that, we first construct $P$ and calculate $P^2$ for HetioNet. After having mapped the relations to primes the outcome would be something similar to the two paths shown in Fig.~\ref{fig:PAM_metapaths}~(a), where the important metapath Compound–\textit{resembles}–Compound–\textit{treats}–Disease (CrCtD) is mapped to the sequence of primes $3,5$, with the relation ``\textit{resembles}'' mapped to $3$ and ``\textit{treats}'' to $5$. This essentially means that this metapath is exhibited in $P^2$ in the form of $P^2[C1, D] = 3*5 = 15$. Hence, the values found in $P^2$ express different kinds of metapaths, and we simply need to look up the matrix $P^2$  to extract them. 

However, in order to find the ones that imply that a compound treats a disease, we can utilize the pairs of compounds and diseases for which we already know that are also directly connected with a ``\textit{treats}'' relation, in the original knowledge graph. This is the same line of thought as the original work, where they use this knowledge as a supervision signal for training their models.

Once again, this is simply translated into looking up the pairs for which $P[C,D] = 5$ (i.e. ``\textit{treats}'') in the 1-hop PAM. Following this procedure, we can extract possibly important metapaths $M$ from the cells $(C, D)$ for which $P^2[C,D] \neq 0$,  and  $P[C,D] = 5$. For example, the metapath $m_1$ corresponding to the CrCtD path of Fig.~\ref{fig:PAM_metapaths}~(a) would be expressed as $m_1 : 15 \Rightarrow 5$, meaning that if we see a compound and a disease connected through the value $15$ in $P^2$ (i.e. the path CrCtD) , it is possible that this compound should be directly connected with the relation $5$ (i.e. ``\textit{treats}'' ) with the disease.

After completing this process and finding all possible metapaths $M$ of length 2 that imply the relation ``\textit{treats}''  we ended up with $55$ distinct such metapaths. These are more than the ones presented in the original work, because the metapaths in $P^k$ essentially express aggregates of paths, due to the summation of products in the values of the cells of $P^2$. For example, the pattern at the top of Fig.~\ref{fig:PAM_metapaths}~(b) would be expressed as $P^2[C1,D] = 3*5 + 3*5 = 30$ and the corresponding metapath generated from this pair of nodes (if $C1$ was also directly connected with $D$ through a relation $5$), would be $m_2 : 30 \Rightarrow 5$. This metapath is treated as a different one in our approach from the metapath $m_1: 15 \Rightarrow 5$ as it expresses a different kind of aggregation of paths, in this case, a compound connected to a disease through 2 distinct CrCtD paths.

This extra layer of complexity provides us with more flexibility in the patterns that are extracted. For example, the metapath at the bottom of Fig.~\ref{fig:PAM_metapaths}~(b) (with the prime $7$ corresponding to the relation ``\textit{paliates}'') cannot be expressed with the patterns extracted by the original authors, because it is the aggregation of different kinds of paths connecting the same start and end node. However, following our process, it is easy to extract the corresponding metapath $m_3 : 36 \Rightarrow 5$. 

Specifically, looking for compound-disease pairs that are connected through this $m_3$ metapath we find 8 such pairs in the KG. Out of them, 4/8 pairs are also directly connected with the relation ``\textit{treats}'' (supporting our intuition that this also may be an important metapath). From the rest 4/8, which are not directly connected with the ``\textit{treats}'' relationship, 2 pairs were
deemed irrelevant according to biomedical experts. However, for the remaining 2 pairs, the bibliography indicates that the compound in question actually treats the disease to which it is connected through $m_3$. Specifically, we found that the compound Hexoprenaline\footnote{https://go.drugbank.com/drugs/DB08957} is  considered a treatment for the disease asthma\cite{pinder1977hexoprenaline} and the compound Fluticasone propionate\footnote{https://go.drugbank.com/drugs/DB00588} can be used with other drugs in treatment for hematologic cancer\cite{williams2015encouraging}. This indicates that the more complex metapaths that are extracted with our methodology can also be used for novel drug efficacy predictions.

Overall, in this section, we showcased the usefulness of PAMs in the task of metapath mining. For a medium-scale heterogeneous knowledge graph, we can have the high-order PAMs readily available very fast and we can easily extract, through the described look-up and matching process, possibly important metapaths that can be used to offer semantic explainability to novel links proposed by a computational system. Finally, it is also worth noting that in the above procedure, no training of any model was required, as opposed to the original work, in order to find the same important metapaths  along with more complex ones.

\section{Conclusions}\label{sec:conclusions}
In this work, we introduced the Prime Adjacency Matrix (PAM) representation for analysing biomedical knowledge graphs. It allows us to represent whole KGs through a single-adjacency matrix losslessly. This enables us to tackle many downstream tasks on the graph, leveraging tools from classical network analysis. We showcased this by providing brief studies on drug repurposing and metapath extraction. In the future, we aim to further experiment with the representation, firstly by verifying the results of the two case studies presented here in a more rigorous evaluation framework. Also, we will experiment with more complex representations extracted from the PAMs and move on to new tasks, such as drug-target-disease predictions. Finally, it would be interesting to use the framework for different granularity graphs as well, such as representing small-molecules and tackling molecule property prediction tasks.

\bibliographystyle{ieeetr}
\bibliography{bibliography}

\end{document}